\documentclass{article}[10pt]
\usepackage{bm}
\usepackage{amsmath}
\usepackage{amsfonts}
\usepackage{amscd} 
\usepackage{graphicx,color}
\usepackage{color}
\def\sumint{\sum\mspace{-25mu}\int}

\begin{document}
\begin{center}
{\bf Relativistic few-body methods}
\end{center}
\begin{center}
{\bf 
W.~N.~Polyzou\\
Department of Physics and Astronomy\\
The University of Iowa\\
Iowa City, IA 52242, USA}
\end{center}

\begin{center}
{Contribution to the \it 21-st International Conference on Few-Body Problems
in Physics}
\end{center}

\begin{itemize}{ 
\item [] I discuss the role of relativistic quantum mechanics in few-body
physics, various formulations of relativistic few-body quantum
mechanics and how they are related. 
}
\end{itemize}  
%
%
\section{Introduction}

The observable degrees of freedom in any experiment are particles.
Particles appear as initial and final states in scattering
experiments.  They are also the elementary building blocks of complex
systems from nuclei to stars.  While interactions do not conserve
particle number, if the $S$-matrix is unitary then all of the reaction
products remain particles.  This observation implies that it must be
possible to reformulate mathematical models of any physical system
directly in terms of particle degrees of freedom.  Since relativistic
invariance is a fundamental symmetry of physical systems, it follows
that it must be possible to represent mathematical models of physical
systems as relativistically invariant quantum theories of particles.

If the underlying theory is a quantum field theory that 
describes interacting particles, it must have a representation as a
relativistically invariant quantum theory of particles.  The most
interesting case is QCD, where the field theory involves
non-observable quark and gluon degrees of freedom, but the theory
ultimately describes reactions where the observable initial and final
states are baryons and mesons.  When expressed in terms of mesons and
baryon degrees of freedom, QCD has the structure of a relativistic
quantum theory of interacting particles.  In the particle
representation the degrees of freedom are experimentally observable, but the
interactions necessarily become complicated.  One important challenge
is to construct the interactions between particles directly from QCD.
If these interactions could be determined, it is likely that the particle
representation of QCD would be a more efficient representation for
calculating scattering observables, since the particle degrees of
freedom are directly related to experiment.  For this reason it also
makes sense model these interactions by combining fundamental principles
with experimental constraints.

The general principles of relativistic quantum theory provide
representation independent constraints on these interactions.  These
constraints are non-trivial and may not be compatible with some
truncations of the field theory.  In addition, these constraints
provide guidance in constructing realistic models of few hadron
systems that remain valid at relativistic energies.

I discuss various representation of relativistic quantum theories of
particles and their relation to each other and to different
representations of quantum field theory.  One of the challenges of
this subject is that there are may different representations.  Some
are more suited to computation while others are more directly related
to an underlying quantum field theory.  The equivalence between
different representations is an important tool for understanding the
structure of realistic models.

\section{Relativistic invariance - general considerations}

Relativistic invariance is a symmetry of quantum theory which requires
that quantum probabilities for equivalent experiments performed in different
inertial frames are identical.  Mathematically this requires that
there is a unitary representation of the Poincar\'e group \cite{Wigner:1939cj}
$U(\Lambda ,a)$ 
\begin{equation}
U(\Lambda_2 ,a_2)U(\Lambda_1 ,a_1) =
U(\Lambda_2\Lambda_1 ,\Lambda_2 a_1 + a_2) 
\label{eq:1}
\end{equation}
(the semidirect product of the Lorentz group with the spacetime
translation group) on the Hilbert space of the quantum theory.  Here
$\Lambda$ is a Lorentz transformation connected to the identity and
$a$ is a space-time translation parameter.

In addition, the ability to test special relativity by
performing measurements on isolated subsystems requires that the
unitary representation of the Poincar\'e group asymptotically
approaches a tensor product of subsystem representations
when the subsystems are asymptotically separated:
\begin{equation}
\lim_{\vert r_{i} - r_j\vert \to \infty} 
U(\Lambda ,a) =  U_{i}(\Lambda ,a) \otimes U_j(\Lambda ,a).
\label{eq:2}
\end{equation}
The smallest of these isolated subsystems are single-particle
subsystems, which are the building blocks of the particle
representation.  The constraints (\ref{eq:2}) are fundamental
constraints on acceptable models or truncations of field theories to a
computable number of degrees of freedom.

\section{Hilbert space representations}

While there are many ways to construct unitary representations of the
Poincar\'e group, the common feature of all of these representations
is that they can always be decomposed into direct integrals of
irreducible representations.  This decomposition is the relativistic
analog of diagonalizing the Hamiltonian in non-relativistic quantum
mechanics.  The diagonalization is more complicated in the
relativistic case because it is necessary to simultaneously
diagonalize the mass and spin Casimir operators.  Both of these
operators involve interactions in relativistic theories that satisfy
(\ref{eq:2}).  

The structure of the physical Hilbert space has a significant effect on
the structure of the dynamical unitary representation of the Poincar\'e group. 
Covariant representations,  which are typical of field theories,  
have Hilbert space inner products with non-trivial kernels,
where the dynamics is encoded in the kernel.  In explicit 
particle representations the dynamics appears in the unitary 
representation of the Poincar\'e group.  The three most common representation 
are discussed below.

\begin{itemize}

\item[1.] Local field theory.  In this representation vacuum expectation 
values of products of fields appear in the kernel the Hilbert space scalar 
product:
\begin{equation}
W_n (x)= W_n (x_1, \cdots, x_n) := \langle 0 \vert \phi(x_1) \cdots \phi(x_n) \vert 0 \rangle 
\label{eq:3}
\end{equation}
\begin{equation}
\langle f \vert g \rangle 
 = \sum_{mn} \int f^*_m(x)W_{m+n} (x,y )  g_n (y) d^{4m}x d^{4n}y .
\label{eq:4}
\end{equation}
\item[2.] Euclidean field theory.  In this representation the 
Euclidean Green functions, 
\begin{equation}
S_n (x) = S_n(x_1 \cdots x_n) = \int D[\phi] d^{-A[\phi]}\phi(x_1) \cdots
\phi(x_n),
\label{eq:5}
\end{equation}
which are formally moments of a Euclidean path integral, appear in  
the kernel of the physical Hilbert space inner product
\cite{Osterwalder:1973dx}
\begin{equation}
\langle f \vert g \rangle 
 = \sum_{mn} \int f^*_m(Rx)S_{m+n} (x;y )  g_n (y) d^{4m}x d^{4n}y
\label{eq:6}
\end{equation}
where $R$ is the Euclidean time reversal operator and the test 
functions $f$ have support for positive relative Euclidean times.  
The collection of
Euclidean Green functions $\{S_n\}$ are called reflection positive if
$\Vert f \Vert = 
\langle f \vert f \rangle$  above is non-negative.

\item[3.] Direct interaction representations.  The Hilbert space is
the direct sum of tensor products of mass $m$ spin $j$ irreducible
representation spaces ${\cal H}_{mj}$:
\begin{equation}                                                               {\cal H} = \oplus (\otimes {\cal H}_{mj}) .                                    
\label{eq:7}
\end{equation}
This is the Hilbert space for a collection of free relativistic
particles.  Explicit dynamical representations of the Poincar\'e group
or Lie algebra are constructed on this space.

\item[4.] Schwinger-Dyson - quasipotential representations.   In this 
representation the dynamical quantities are time-ordered Green functions 
\begin{equation}
G_n(x_1, \cdots ,x_n) = \langle 0 \vert T(\phi(x_1) \cdots \phi(x_n)) 
\vert 0 \rangle 
\label{eq:8}
\end{equation}
\item [] where $T$ is the time ordering operator. 

\item [] While these functions do not have a direct relation to 
a Hilbert space inner product,  they can be used to calculate
matrix elements of any operator between physical particle states
\cite{Huang:1974cd}.  In principle both the time-ordered Green functions, 
$G_n$, and the Wightman functions, $W_n$, are related to the Euclidean 
Green functions, $S_n$, by different analytic continuations. It
follows that both have the same dynamical content.
 
\item [] Quasipotential methods replace the equations for 
the time-ordered Green functions by equivalent equations for
quasipotentials and constrained time ordered Green functions.
The simplest example is the  
Schwinger-Dyson equation for the four point function, $G_4$, which is the
Bethe-Salpeter equation:
\begin{equation}
G_4 = G_0 + G_0 K G_4 \qquad G_0 = {(G_2 G_2)}_{symmetrized}
\label{eq:9}
\end{equation}
Here $K$ is the Bethe Salpeter kernel which is formally defined by 
(\ref{eq:9}).  The four point function 
can be expressed in terms of the Bethe-Salpeter $T$ by
\begin{equation}
G_4 = G_0 + G_0 T G_0 
\qquad \mbox{where} \qquad T = K + KG_0 T .
\label{eq:10}
\end{equation}
In the quasi-potential formulation $G_0$ is replaced by a simpler $g_0$ 
with the same singularities.  The equation for $T$ is replaced 
by the pair of equations: 
\begin{equation}
G_0 = g_0 + \Delta g
\label{eq:11}
\end{equation}
\begin{equation}
T = U + U g_0 T
\qquad
U = K + K \Delta g U
\label{eq:12}
\end{equation}
\item[] Quasipotential methods provide an alternative set of equations to 
calculate the time-ordered Green functions.  
\end{itemize}

\section{Unitary representations of  the 
Poincar\'e group}

In the case of quantum field theory, the covariance condition
\begin{equation}
W_n (x_1, \cdots, x_n) = W_n (\Lambda x_1 +a , \cdots, \Lambda x_n+a) 
\label{eq:13}
\end{equation}
implies that the mapping
\begin{equation}
f_k (x_1, \cdots, x_k) = f_k (\Lambda x_1 +a , \cdots, \Lambda x_k+a) 
\label{eq:14}
\end{equation}
preserves the inner product (\ref{eq:4}) and is thus defines a
unitary representation of the Poincar\'e group.

In the Euclidean case the Euclidean time reversal operator $R$ in
(\ref{eq:4}) converts a unitary representation of the Euclidean group to a
representation of a subgroup of the complex Poincar\'e group.  

The Hamiltonian and rotationless boost generators in this representation are 
\begin{equation}
H g_k(x) = \sum_j {\partial g_k \over \partial \tau_j} (x)
\qquad
K^i g_k(x) = \sum \tau_j {\partial g_k \over \partial x_{ij}} (x)-
x_{ij} {\partial g \over \partial \tau_j} (x).
\label{eq:15}
\end{equation}
These are self-adjoint with respect to the inner product (\ref{eq:6}) and 
along with the generators of translations and rotations satisfy 
the commutation relations of the Poincar\'e Lie algebra.  

For particle representations the problem is to construct a set of
Poincar\'e generators on the particle Hilbert space.  Once the mass
and spin operators are simultanelously diagonailzed the structure of
the irreducible representations is fixed by group representation
theory.  On each irreducible subspace with vectors labeled by three
momentum and the projection of canonical spin on an axis the
transformation, has the form
\begin{equation}
U (\Lambda ,0) \vert (m,j) p, \sigma \rangle_c  = \sum
\vert (m,j)\Lambda p , \sigma' \rangle_c
\sqrt{{\omega_m (\pmb{\Lambda}p) \over \omega_m(\mathbf{p})}}
D^j_{\sigma'\sigma}[R_w(\Lambda,p)]  
\label{eq:13}
\end{equation}
where $R_w(\Lambda ,p)$ is a Wigner rotation and
$\omega_m(\mathbf{p})$ is the energy of a particle of mass $m$ and
momentum $\mathbf{p}$.  Dirac suggested a number of simplifying
representations\cite{Dirac:1949cp}, called the instant, point and
front-form of dynamics.  They are distinguished by the choice of
subgroup of the Poincar\'e group that is free of interactions.  The
different choices, called Dirac's forms of dynamics
are all related by $S$-matrix preserving unitary transformations
\cite{Sokolov:1977im}.

The connection between all three representations is easily seen 
for a single spinless free particle of mass $m$.  In this case 
the two point functions $W_2(x-y)$ and $S_2(x-y)$ have well known 
forms.  The wave function in the Minkowski and Euclidean covariant 
representations are related to the wave function in the particle 
representation by
\[
f_p (\mathbf{p}) =
{1 \over (2 \pi)^{3/2}} \int
{e^{- i \mathbf{p} \cdot \mathbf{x} + i \omega_{\mu}(\mathbf{p})t}
\over \sqrt{2 \omega_{\mu} (\mathbf{p})}}
f_m (t, \mathbf{x}) d^4 x =
\]
\begin{equation}
{1 \over (2 \pi)^{3/2}} \int
{e^{- i \mathbf{p} \cdot \mathbf{x} - \omega_{\mu}(\mathbf{p})\tau}
\over \sqrt{2 \omega_{\mu} (\mathbf{p})}}
f_e (\tau, \mathbf{x}) d^4 x .
\label{eq:14}
\end{equation}

The simplest was to understand the origin of these relations is 
to observe that the
kernel of the inner product (\ref{eq:2}) is a positive 
hermetian quadratic form.
If it was a matrix it could be written as the square of the 
positive square root of the matrix.  These square roots could then 
be absorbed into the wave functions, thus removing the kernel 
of the quadratic form.   What is actually done is to insert a 
complete set of intermediate particle states between the 
fields in the initial and final parts of the scalar product.
This leads to an expression like
\[
\langle f \vert g \rangle = 
\sumint f^*(x_1, \cdots ,x_n) \langle \phi(x_1) \cdots \phi(x_n)\vert (m,l,d)\mathbf{p},\mu_l 
\rangle d \mathbf{p} \times
\]
\begin{equation}
\langle (m,l,d)\mathbf{p},\mu_l  \vert 
\phi(y_m) \cdots \phi(y_1) \vert 0 \rangle
g (y_1, \cdots ,y_m)d^{4m}y d^{4n}x 
\label{eq:15}
\end{equation}
where the $d$ represents degeneracy labels.  The quantities 
\begin{equation}
f((m,l) d,\mathbf{p},\mu_l):= 
\int \langle (m,l,d)\mathbf{p},\mu_l \vert   
\phi(y_m) \cdots \phi(y_1) \vert 0 \rangle
f^*(y_1, \cdots ,y_m) d^{4m} y
\end{equation}
are wave functions of the particle representation.
The amplitudes 
\[
\langle \phi(x_1) \cdots \phi(x_n)\vert (m,l,d)\mathbf{p},\mu_l 
\rangle
\]
intertwine finite dimensional representation of the 
the Lorentz group with unitary irreducible representations of the
Poincar\'e group:
\[
\langle \phi(\Lambda x_1) \cdots \phi(\Lambda x_n)\vert (m,l,d)\mathbf{p},\mu_l
\rangle =
\]
\begin{equation}
\langle \phi(x_1) \cdots \phi(x_n)\vert (m,l,d)\pmb{\Lambda}p,\nu_l
\rangle D^{l}_{\nu_l \mu_l}[R_w(\Lambda ,p)] 
\end{equation}
where $R_w(\Lambda ,p)$ is a Wigner rotation.  This is a generalization 
or standard intertwining properties of the $u(p)$ and $v(p)$ Dirac spinors.

\section{Applications}

Relativistic quantum mechanics is a powerful framework for
investigation systems of light quarks.  Relativistic quantum
mechanical models have been used to look at spectral properties,
lifetimes, and electroweak observables.  The have also been used in
few-hadron systems to study bound state, scattering and
electromagnetic observables.

One illustrative example is using $p-d$ scattering in the few GeV
range.  At this scale one expects that one might be sensitive to
sub-nucleon degrees of freedom that cannot be easily modeled in a
meson-exchange picture.  In examining data it is important to be able
to separate relativistic effects from potential sub-nucleon effects.
Figure one shows the result of a relativistic direct interaction p-d
calculation \cite{Lin:2007kg} scattering calculations with a Malfliet
Tjon interaction at 500 MeV.  The plots show the breakup cross section
as a function of the energy of one of the protons when the final two
protons make equal angles with the beam line.  The solid curves are
for the relativistic calculation, the long dashes are for the
non-relativistic calculation.  The relativistic and non-relativistic
calculations are constrained to give the same two-body CM cross
sections.  The data is from \cite{Punjabi:1988hn}.  This observable
illustrates as strong relativistic effects, as non-relativistic and
relativistic curves change places as the angle increases.

Figure 2 shows the elastic p-d scattering cross section at 135 and 250
MeV\cite{PhysRevC.83.044001} \cite{PhysRevC.88.069904}(erratum).  The
dash-dot curve is a full relativistic calculation with realistic two (CD
Bonn) and three-nucleon (TM99) forces, the long dashed curve is a
relativistic calculation with only two-nucleon forces, the dotted
curve is a non-relativistic calculation with two and three-nucleon
forces, and the solid curve is a non-relativistic calculation with two
nucleon forces.  The 135 MeV calculations show no significant
relativistic effects.  Agreement with data \cite{Sekiguchi:2002sf} for
scattering angles larger than 60 degrees is explained by the inclusion
a three-nucleon force.  At 250 MeV there are enhancements at back
angles due to both relativistic effects and three-nucleon force
effect, but the combined effect of both of these are almost an order
of magnitude too small to explain the discrepancy with the
data\cite{PhysRevC.66.044002}\cite{PhysRevC.76.014004}(nd).  
Back angles correspond to harder scattering and is were one 
might expect to begin seeing discrepancies from meson exchange models.
Since the energy is still below the pion production threshold, this
is a clear indication of missing physics in standard realistic meson
exchange nuclear forces.

The application of quantum relativistic methods to few-nucleon
problems at the Few Gev scale will bring new powerful tools to study
nuclear reactions in the energy region where one expects to see a
transition from the dominance by hadronic degrees of freedom to one
dominated by QCD degrees of freedom.

I would like to acknowledge the may scientists with whom I have had
the pleasure of collaborating with on various problems related to
relativistic quantum mechanics.  These include W. Gl\"ockle (Bochum),
T. Allen, P. L. Chung, F. Coester, H. C. Jean, W. Klink, P. Kopp,
M. Herrmann, S. Kuthini Kunhammed, G. L. Payne, Y. Huang (Iowa),
J. Golak, R. Skib\'inski, K. Topolnicki, H. Wita{\l}a (Jagiellonian U)
H. Kamada (Kyushu), B. Keister (NSF), Ch. Elster, T. Lin, M. Hadizadeh
(Ohio).  This research was supported by U.S. DOE contract
No. DE-FG02-86ER40286 and NSF contract No.  NSF-PHY-1005501 .

\begin{figure}
\begin{minipage}[t]{.48\linewidth}
\centering
\includegraphics[scale=.5]{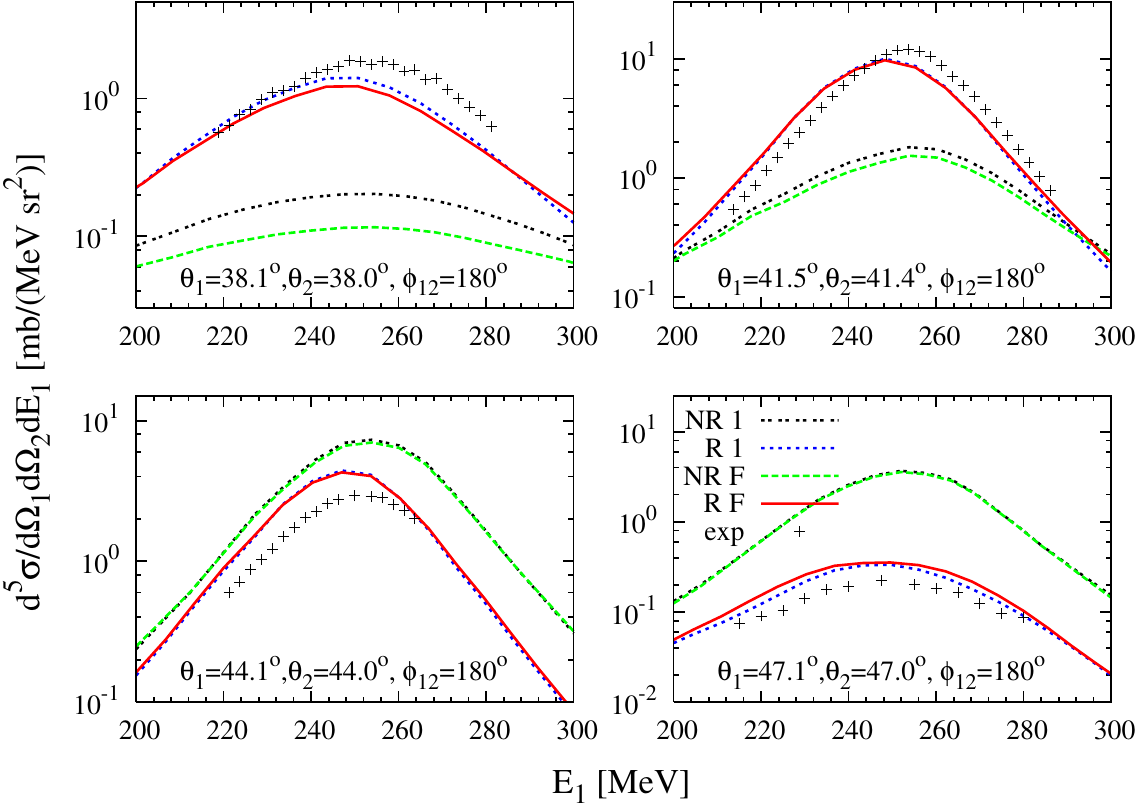}
\caption{breakup - symmetric outgoing protons}
\label{fig:1}
\end{minipage}
\begin{minipage}[t]{.35\linewidth}
\centering
\includegraphics[scale=.35]{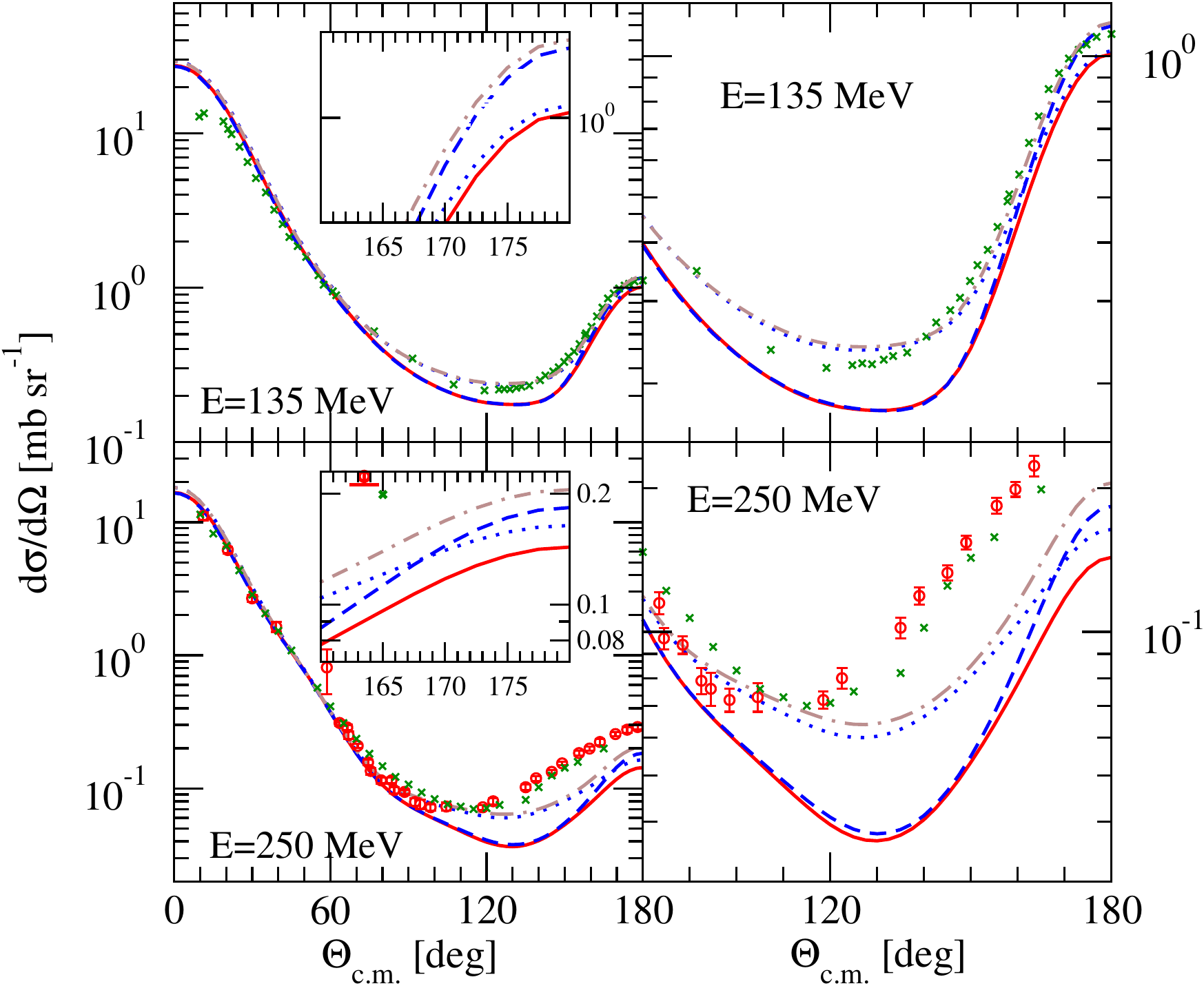}
\caption{elastic pd scattering}
\label{fig:2}
\end{minipage}
\end{figure}

\end{document}